# Molecular Dynamics Analysis of Graphene-Based Nanoelectromechanical Switch

Eunae Lee[a] and Jeong Won Kang[a,b]

[a] Department of IT Convergence, Korea National University of Transportation, Chungju 380-702, Republic of Korea
[b] Graduate School of Transportation, Korea National University of Transportation, Uiwang-si, Gyeonggi-do 437-763, Republic of Korea

Here we present graphene-based nanoelectromechanical switch with the vertical carbon nanotube electrode via classical molecular dynamics simulations. The carbon nanotube is grown in the center of the square hole and the graphene covers on the hole. The potential difference between the bottom of the hole and the graphene is applied to deflect the graphene. By performing classical molecular dynamic simulations, we investigate the electromechanical properties of graphene-based nanoelectro-mechanical switch with carbon nanotube electrode, which can be switched by the externally applied force. This simulation work explicitly demonstrated that such devices are applicable to nanoscale sensors and quantum computing, as well as ultra-fast-response switching devices.

## Introduction

Graphene (1), a single-atom sheet, has been considered as the most promising material for making future nanoelectromechanical systems (NEMSs) (2) as well as purely electrical switching with graphene transistors (3–5). Graphene-based devices have advantages in scaled-up device fabrication due to the recent progress in large area graphene growth and lithographic patterning of graphene nanostructures (6–10). Graphene based nanoelectromechanical (NEM) switches have been fabricated and explored primarily 2-terminal, one-off laboratory scale demonstrations (11–16). These switches operated by deflecting a suspended graphene membrane with a source-drain voltage and measuring the current once contact was made, but little success has been achieved in terms of repeatable switching (15, 16) because these graphene contact switches reported to date are primarily doubly clamped beams that suffer reliability problems due to tears on open edges and/or irreversible stiction of graphene (11, 12).

Liu et al. (17) fabricated and characterized a large array of circularly clamped graphene-based NEM switches, which can work with either 2-terminal or 3-terminal electromechanical switching. They provided the low actuation voltage and improved mechanical integrity with a novel design, which reduced the contact area thereby reducing stiction problems. 3-terminal NEMS switch using a third electrode to apply an actuation voltage ($V_g$) independent of the $V_{sd}$, provides further advantages such as greater operational flexibility, lower power consumption, and higher level of integration and system functionality compared to 2-terminal devices (18–20).

Here, we present the schematics of 3-terminal graphene-based NEM switch with vertically grown carbon nanotube (CNT) (21) electrode. By performing classical molecular dynamic (MD) simulations, we investigate the electromechanical properties of graphene-based NEM switch with CNT electrode, which can be switched by the externally applied force. This graphene-based NEM switch can be applied to NEMS memory or switching devices.

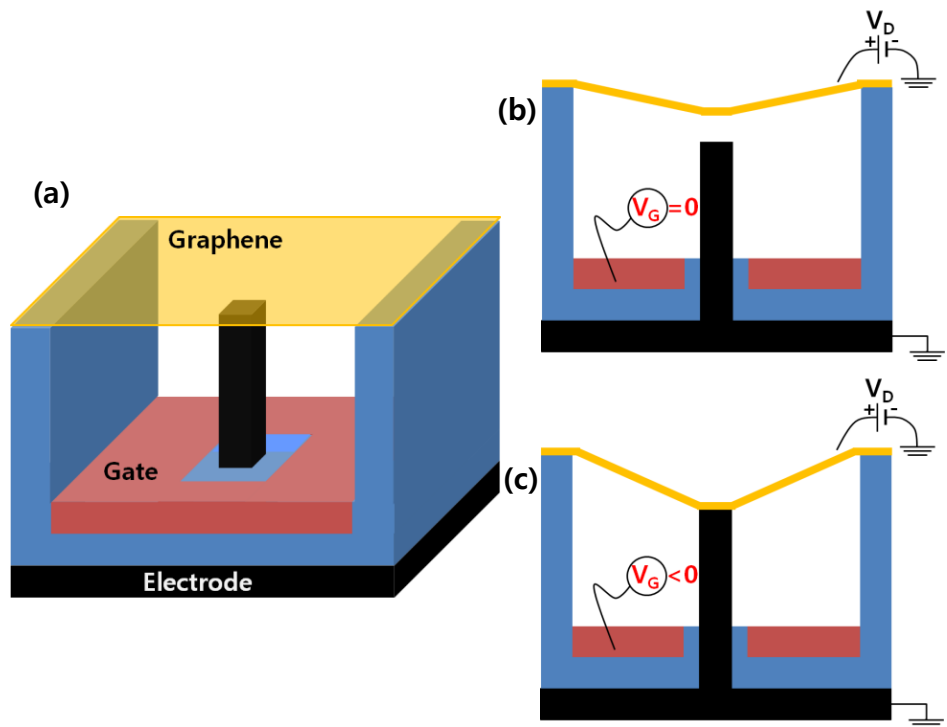

Figure 1. (a) Model schematics for the proposed 3-terminal graphene-based NEM switch with vertically-grown CNT electrode. (b) OFF state. (c) ON state.

## Design and Methods

Figure 1(a) shows the model schematics for the proposed 3-terminal graphene-based NEM switch with vertically grown CNT electrode. The CNT is grown in the center of the square hole and the graphene covers on the hole. The potential difference between the bottom of the hole and the graphene is applied to deflect the graphene. Initially, when the potential difference is zero, the electric current between the CNT and the graphene is also zero; i.e. 'OFF' state as shown in Figure 1(b). When the potential difference increases, the electrostatic charges are induced on both the graphene and the bottom electrode. The electrostatic charges give rise to an electrostatic force, which deflects the graphene, and then the graphene can begin to deflect. When the graphene is deflected to the CNT tip, in addition to electrostatic forces, depending on the gap between the graphene and the bottom electrode, the long range interatomic interactions such as van der Waals interactions also act on the graphene and affect the deflection of the graphene. The electrostatic and the interatomic forces induce the graphene to bend toward the CNT tip. Counteracting the electrostatic and interatomic forces are elastic forces, which try to restore the graphene to its original straight position. Finally, the graphene comes into contact with the CNT tip, and this situation can be detected by sensing electric current between the graphene and the CNT tip as 'ON' sate as shown in Figure 1(c).

In this work, we perform classical MD simulations to measure the switching motion of the graphene-based NEM switch with vertical CNT electrode. We use in-house MD code that has been used in our previous works (22–24) applying the velocity Verlet algorithm, a Gunsteren–Berendsen thermostat to control the temperature, and neighbor lists to improve the computing performance. The MD time step was $5\times10^{-4}$ ps. We assigned the initial atomic velocities with the Maxwell distribution, and the magnitudes were adjusted in order to fit the temperature of the system. In all the MD simulations, the temperature was set to 1 K. In the classical MD simulations, the graphene with $11.5 \times 11.6$ nm$^2$ is composed of 5320 carbon atoms, and the capped (5, 5) CNT is composed of 250 carbon atoms. The interaction between the carbon atoms that form the covalent bonds for graphene were described with the Tersoff-Brenner potential (25, 26). The long-range interactions for carbon between the graphene and the CNT were characterized according to the Lennard-Jones 12–6 (LJ12–6) potential with parameters given by Mao et al. (27). In this paper, the parameters for the LJ12–6 potential were $\varepsilon = 0.0042$ eV and $\sigma = 0.337$ nm. The cutoff distance for the LJ12-6 potential was of 2 nm.

## Results and Discussion

Figure 2 shows the maximum deflection of the graphene as a function of the potential difference between the graphene and the bottom electrode. The maximum deflection of the graphene almost linearly increased with increasing the potential difference when $V_g < 30$ V. However, when $V_g \geq 30$ V, the maximum deflection of the graphene slowly increased with increasing the potential difference.

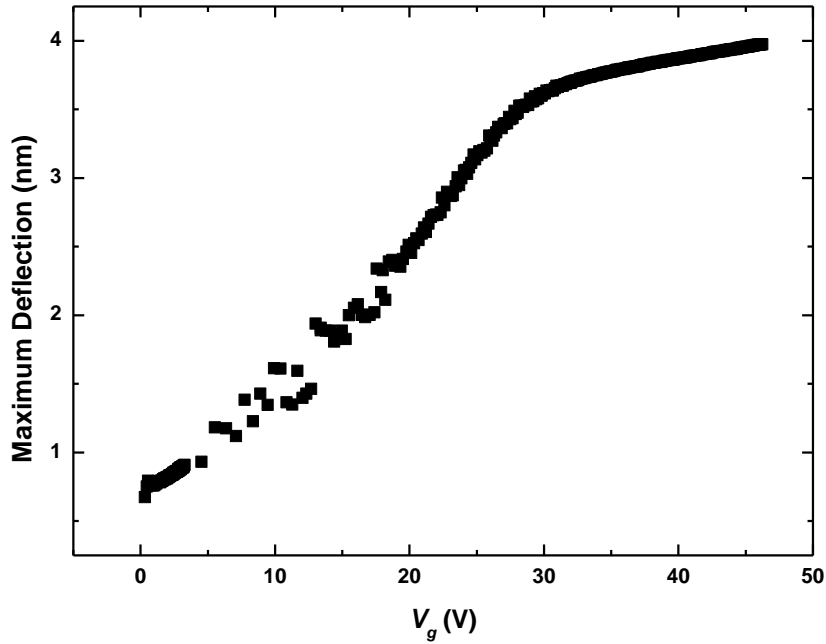

Figure 2. Maximum deflection of graphene as a function of the potential difference between graphene and the bottom electrode

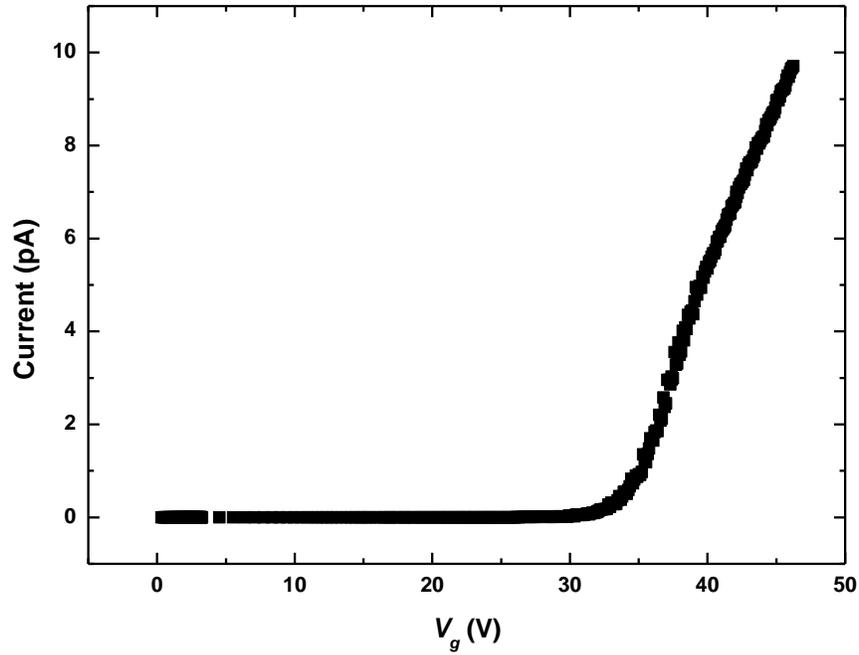

Figure 3. Tunneling current as a function of the potential difference between graphene and the bottom electrode.

We calculated the electric current of the CNT tip using the tunneling current approximation in scanning tunneling microscopy (STM) (28, 29). Figure 3 shows the electric current as a function of the potential difference. The electric current abruptly increased with increasing the potential difference when $V_g \geq 35$ V. Hence, the graphene-based NEM switch with vertical CNT tip has a great potential to be switching device. In this work, the switching voltage was about 35 V. However, this switching voltage can be reduced by using the larger size of the hole. The tunneling currents can be increased by using the metal tips with low work functions.

Such results introduce that as the potential difference increases, the graphene can be deflected more and more and finally can contact with the CNT. Then, the electric current can abruptly increase. However, as the potential difference decreases, the elastic restoring force becomes higher than the electrostatic and the interatomic forces and then, the graphene separates out from the CNT tip, then backs up it, and finally the graphene comes into 'OFF' state. Therefore, the graphene-NEM switch can alternate between the ON and the OFF states by adjusting the potential difference between the graphene and the bottom electrodes.

The MD simulation in this work was performed at an extremely low temperature, very different from the 'ordinary' conditions of experiments or measurements, so the dynamic features of the graphene NEM switch are presented in an ideal situation. This simulation work explicitly demonstrated that such devices are applicable to nanoscale sensors and quantum computing, as well as ultra-fast-response switching devices.

## Conclusion

In conclusion, we presented the simple schematics of the three-terminal nanoelectromechanical switch using the vertical CNT tip and planar graphene and investigated their operation dynamics via classical molecular dynamics simulations combined with classical electrostatic theory. The key operations of the proposed graphene-based nanoelectromechanical switch are based on two issues: first, the low mass density, the large area, and the flexible deflection of the graphene and second, narrow and sharp of the carbon nanotube tip. The electromechanical dynamics of the graphene switch are well balanced by five forces as follows: the capacitive force between the bottom electrode and the graphene, the van der Waals force between the bottom electrode and the graphene, the capacitive force between the carbon nanotube tip and the graphene, the van der Waals force between the carbon nanotube tip electrode and the graphene, and the elastic force of the graphene.

## Acknowledgments

This research was supported by the MSIP (Ministry of Science, ICT and Future Planning), Korea, under the C-ITRC (Convergence Information Technology Research Center) (IITP-2016-H8601-16-1008) supervised by the IITP (Institute for Information & communications Technology Promotion), and the Ministry of Education (MOE) and National Research Foundation of Korea (NRF) through the Human Resource Training Project for Regional Innovation (2014H1C1A1066414).